\documentclass[onecolumn]{autart}    
\usepackage{amsmath, amssymb}
\usepackage{bm}                         
\usepackage[utf8]{inputenc}
\usepackage{graphicx}
\usepackage[version=4]{mhchem}
\usepackage{siunitx}
\usepackage{cite}
\usepackage{longtable,tabularx}
\usepackage{graphicx}          
\usepackage{booktabs}
\usepackage{multirow}

\usepackage{graphicx}
\usepackage{subfigure}
\usepackage{subcaption}
 \newtheorem{theorem}{Theorem}

\newtheorem{corollary}{Corollary}
\newtheorem{remark}{Remark}

\usepackage{epstopdf}

\usepackage{graphicx} 
\usepackage{caption}  

\newcommand{\roa}{\mathcal{R}}
\newcommand{\roahat}{\widehat{\roa}}
\newcommand{\ehat}{\widehat{\mathcal{E}}}
\newcommand{\trans}{^\mathsf{T}}

\graphicspath{{./figures/}}

\begin{document}

\begin{frontmatter}
                                           
\title{Local Stability and Stabilization of\\ Quadratic-Bilinear Systems using Petersen's Lemma}

\thanks[footnoteinfo]{This material is based upon work supported by the Air Force Office of Scientific Research under award number FA9550-22-1-0004 and the National Science Foundation under grant number CBET-1943988. The material in this paper was not presented at any conference.}

\author[1]{Amir Enayati Kafshgarkolaei}\ead{enaya004@umn.edu}\ and    
\author[1]{Maziar S. Hemati}\ead{mhemati@umn.edu}               

\address[1]{Aerospace Engineering and Mechanics, University of Minnesota, Minneapolis, USA 55455}

\begin{keyword}                           
Region of Attraction \sep Quadratic Bilinear Systems \sep Petersen's Lemma \sep Linear Matrix Inequality
\end{keyword}                             

\begin{abstract}             
Quadratic-bilinear (QB) systems arise in many areas of science and engineering.
In this paper, we present a scalable approach for designing locally stabilizing state-feedback control laws and certifying the local stability of QB systems.
Sufficient conditions are established for local stability and stabilization based on quadratic Lyapunov functions, which also provide ellipsoidal inner-estimates for the region of attraction and region of stabilizability of an equilibrium point.
Our formulation exploits Petersen’s Lemma to convert the problem of certifying the sign-definiteness of the Lyapunov condition into a line search over a single scalar parameter.
The resulting linear matrix inequality~(LMI) conditions scale quadratically with the state dimension for both stability analysis and control synthesis, thus enabling analysis and control of QB systems with hundreds of state variables without resorting to specialized implementations.
We demonstrate the approach on three benchmark problems from the existing literature.
In all cases, we find our formulation yields comparable approximations of stability domains as determined by other established tools that are otherwise restricted to systems with up to tens of state variables.

\end{abstract}

\end{frontmatter}

\section{Introduction}

Quadratic-bilinear (QB) systems are nonlinear dynamical systems for which the dynamics are quadratic functions of the state and bilinear functions of the state and input.
When no inputs are present, QB systems reduce to quadratic systems.
QB systems arise in many areas of the natural, social, and applied sciences, including aerospace engineering~\cite{faruqi}, fluid dynamics~\cite{Moehlis2004}, chemical processes~\cite{Bequette1998}, atmospheric science~\cite{Lorenz1963}, ecology~\cite{berryman1992}, electrical circuits~\cite{karachalios2022}, epidemiolgy~\cite{MarinovMarinova2022}, and robotics~\cite{SciaviccoSiciliano2012}.
QB system models also arise naturally in second-order approximations of nonlinear systems governed by analytic functions.
As such, an ability to certify the stability of QB systems and to reliably stabilize QB systems with feedback control stands to benefit a variety of fields.

Tailored methods for stability analysis and control of QB systems have been developed by exploiting the underlying problem structure associated with QB systems.
However,  as noted in~\cite{Kramer21}, prevailing methods for analysis and control of QB systems tend to be limited to low-order systems with state dimension $n\lesssim\mathcal{O}(10)$ due to computational challenges; this is problematic because even reduced-order models of QB systems arising in many applications commonly consist of $n\sim\mathcal{O}(10)$--$\mathcal{O}(100)$ state variables~\cite{Kramer21,bennerACM2024}.
For instance, it is possible to certify stability of a QB system within a polytope, but problem size scales exponentially with state dimension, thus limiting the approach to relatively small systems~\cite{amato2006region}.
Related methods for stabilization of QB systems have also been proposed, but inherit the same exponential increase in problem size with state dimension~\cite{amato2007state}.

Methods for certifying stability within simpler geometries (e.g.~ellipsoids) have also been proposed.
If a valid quadratic Lyapunov function is already known, then it is possible to determine an optimal ellipsoid
in which stability can be certified~\cite{Tesi1994,tesi1994TAC}.
Accordingly, an analytical approach was proposed in~\cite{Kramer21} for constructing a valid Lyapunov function upon which the optimization of~\cite{Tesi1994,tesi1994TAC} can be applied, making it possible to estimate stability domains for QB systems with $n\sim\mathcal{O}(10)$ state variables.
However, the resulting stability estimates will be conservative in general because the quality of the optimal stability estimate is specific to a given Lyapunov function, which is itself not included as an optimization variable in this setting.
Recently, so-called \emph{quadratic-constraint-based methods} have been used to certify stability in QB systems with state dimension $n\sim\mathcal{O}(100)$.
Quadratic-constraint-based methods consider a Lur'e decomposition of the QB system into a feedback interconnection between the linear dynamics and the nonlinear terms.
In this setting, stability analysis is conducted using dissipation inequalities with quadratic constraints used to bound the influence of the nonlinear terms on the linear dynamics~\cite{KalurEtAl21, KalurSeilerHemati21}.
The approach ultimately amounts to analyzing computationally tractable linear matrix inequality~(LMI) conditions.
Recent advances have further exploited the underlying problem structure to refine the quadratic constraints used to bound the nonlinearity, thus reducing conservatism~\cite{LiaoSeilerHemati22,Liu2020,TosoEtAl22}.

In this paper, we take advantage of Petersen's Lemma~\cite{Petersen87,Khlebnikov2008} to derive tractable LMI conditions for stability analysis and stabilizing-controller synthesis for QB systems.
Petersen's Lemma~\cite{Petersen87}---to be discussed in Section~\ref{sec:background}---has recently been used to certify stability and design stabilizing feedback control laws for bilinear systems~\cite{Khlebnikov2015,Khlebnikov2017}, and also in data-driven control for nonlinear systems~\cite{BisoffiDePersisTesi22}.
Here, we exploit the specific structure of QB systems and apply Petersen's Lemma to derive 
our main synthesis result: an LMI condition that can be used to compute locally stabilizing state feedback control laws for QB systems (see  Theorem~\ref{thm:main_synthesis} in Section~\ref{sec:synthesis}).
Our main analysis result follows naturally from the synthesis result and provides an LMI condition from which to compute ellipsoidal approximations of the ROA in quadratic systems (see Corollary~\ref{thm:main_analysis} in Section~\ref{sec:synthesis}).
We provide several illustrative examples in Section~\ref{sec:analysis_ex} to highlight two key features of our approach: (1)~the approach scales to systems with state dimension $n\sim\mathcal{O}(100)$, and (2)~conservatism is comparable to existing methods that do not scale beyond $n\lesssim\mathcal{O}(10)$, and reduced relative to other methods that do scale up to~$n\sim\mathcal{O}(100)$.

\section{Background and Preliminaries}
\label{sec:background}

We begin with necessary background material: Section 2.1
introduces quadratic-bilinear (QB) systems, and Section 2.2 introduces local stability of an equilibrium point and presents a statement of Petersen's Lemma.

\subsection{Quadratic-Bilinear Systems}

Consider the quadratic-bilinear~(QB) system 
\begin{equation}
    \dot{x} = A x + H(x \otimes x) + \sum_{j=1}^{m} D_j x u_j + B u,
    \label{eq:qb}
\end{equation}
where \( x = x(t) \in \mathbb{R}^n \) is the state, \( u = u(t) \in \mathbb{R}^m \) is the input, $t\in\mathbb{R}$ is time, and \( \otimes \) denotes the standard Kronecker product.
The system coefficients are contained in the matrices \( A \in \mathbb{R}^{n \times n} \) and \( B \in \mathbb{R}^{n \times m} \) for linear terms, \( H=\begin{bmatrix}H_1&\cdots&H_n\end{bmatrix} \in \mathbb{R}^{n \times n^2} \) with $H_i\in\mathbb{R}^{n\times n}$ for quadratic terms, and \( D_j \in \mathbb{R}^{n \times n} \) (for \( j = 1, 2, \ldots, m \)) for bilinear terms. 
We take \( H \) to be symmetric such that
\begin{equation}
    H(x_1 \otimes x_2) = H(x_2 \otimes x_1) \qquad \forall x_1,x_2 \in \mathbb{R}^n,
    \label{eq:Hsymm}
\end{equation} which is without loss of generality and does not affect the system dynamics (see, e.g.,~\cite{Benner15}).
Note that the quadratic terms can equivalently be expressed as
\begin{equation}
H(x \otimes x) = \sum_{i=1}^n H_i x \, e_i^T x,
\label{eq:expandquad}
\end{equation}
where \( e_i \) represents the \(i^\mathrm{th}\) standard basis vector for \( \mathbb{R}^n \).

In the autonomous case with $u=0$, the QB system~\eqref{eq:qb} reduces to a quadratic system
\begin{equation}
    \dot{x} = A x + H(x \otimes x).
    \label{eq:quad}
\end{equation}
In the case of a constant input or state feedback control $u=Kx$ with static gain $K\in\mathbb{R}^{m\times n}$, \eqref{eq:qb} reduces to \eqref{eq:quad} with appropriate redefinitions of $A$ and $H$.

\subsection{Local Stability of an Equilibrium Point}

Stability is an inherent property of a given equilibrium point.
An equilibrium point \( x=x_e \) of the quadratic system~\eqref{eq:quad} satisfies
\begin{equation}
    0 = A x_e + H(x_e \otimes x_e).
    \label{eq:nonzero_equilibrium}
\end{equation}
We will take $ x_e=0$ without loss of generality. 
Note that if \( x_e \neq 0 \), then we can perform a variable shift to obtain an equivalent representation of the dynamics in which the equilibrium is at the origin.
To see this, introduce a new variable $z = x - x_e$ such that $z=z_e=0$ when $x=x_e$.
Upon substituting \( x = z + x_e \) into~\eqref{eq:quad} and noting that $x_e$ satisfies~\eqref{eq:nonzero_equilibrium} and that $H$ is symmetric according to~\eqref{eq:Hsymm}, we obtain the following quadratic system with the desired equilibrium point at the origin:
\[\dot{z} = \left[A + 2 H(I_n \otimes x_e)\right] z + H(z \otimes z),\] 
where $I_n\in\mathbb{R}^{n\times n}$ is the identity matrix.

If $A$ is Hurwitz, then the zero equilibrium point of the quadratic system~\eqref{eq:quad} will be asymptotically stable~\cite{Khalil02}.
To ease the ensuing discussions, we will abuse terminology and refer to stability of the system~\eqref{eq:quad} wherein we are referring to stability of the equilibrium point of~\eqref{eq:quad} that is located at the origin. 
Let \( \phi(t, x(0)) \) denote the solution of~\eqref{eq:quad} at time \( t \) from the initial condition \( x(0) \). 
The  region of attraction~(ROA)---also known as the domain of attraction, basin of attraction, or region of asymptotic stability---is the region of state-space for which any trajectory with an initial condition within the ROA converges to the equilibrium point itself.
Formally, the ROA for the equilibrium point $x_e=0$ is defined as
\begin{equation}\label{eq:ROA}
\mathcal{R} := \left\{ x(0) \in \mathbb{R}^n \; : \; \lim_{t \to \infty} \phi(t, x(0)) = 0 \right\}.
\end{equation}
If \( \mathcal{R} = \mathbb{R}^n \), then the equilibrium point \( x_e = 0 \) is globally asymptotically stable.
Otherwise, the equilibrium point will be locally asymptotically stable and the closed set $\mathcal{R}$ can take on some complicated shape within the state-space.
A closely related concept for systems with control inputs is that of the \emph{region of stabilizability~(ROS)}---i.e.,~the region of state-space for which there exists a stabilizing controller to asymptotically steer the system from any state within this region to the origin.
The stabilizing controller need not be the same for all points within the ROS.

An inner approximation of the ROA $\roahat\subseteq\mathcal{R}$ using simple geometric shapes such as ellipsoids~\cite{Kramer21} or polyhedra~\cite{amato2006region} is often sought in order to facilitate analysis.
Ellipsoidal estimates for the ROA of a controlled (closed-loop) system serve as inner-approximations for the ROS, and are sometimes referred to as \emph{stabilizability ellipsoids}~\cite{Khlebnikov2015,Khlebnikov2017}.
The following lemma is commonly employed in determining ROA estimates, and by extension ROS estimates as well.
\begin{lem}[\cite{Khalil02}]
\label{lemma:roahat}
    A given closed set $\roahat\subseteq\mathbb{R}^n$, with $0\in\widehat{\mathcal{R}}$, is an estimate of the region of attraction~(ROA) of the nonlinear system~\eqref{eq:quad} if
    \begin{enumerate}
        \item $\widehat{\mathcal{R}}$ is an invariant set, and 
        \item there exists a Lyapunov function $V(x)$ such that
        \begin{enumerate}
            \item $V(x)>0$  $\forall x\in\roahat\backslash\{0\}$ with $V(0)=0$ and
            \item $\dot{V}(x)<0$ $\forall x\in\roahat\backslash\{0\}$ with $\dot{V}(0)=0$.
        \end{enumerate}
    \end{enumerate}
\end{lem}

In Section~\ref{sec:synthesis}, we will make use of Lemma~\ref{lemma:roahat} to compute ellipsoidal estimates for the ROA of~\eqref{eq:quad}, 
and to synthesize state feedback control laws $u=Kx$ that guarantee stabilization of the QB system~\eqref{eq:qb} within an ellipsoid centered about the origin (i.e.,~within a stabilizability ellipsoid). 
To this end, we introduce Petersen's Lemma, which will be central to our formulations. 
%

\begin{lem}[Petersen's Lemma \cite{Petersen87,Khlebnikov2008}]
\label{lemma:petersen}
Let $G = G^\top$, $M \neq 0$, and $N \neq 0$ be matrices of appropriate dimensions. Then, for all matrices $\Delta \in \mathbb{R}^{p \times q}$ satisfying $\|\Delta\| \leq 1$, the inequality
\begin{equation}
G + M \Delta N + N^\top \Delta^\top M^\top \prec 0,\label{eq:ineqpetersen}
\end{equation}
holds if and only if there exists a scalar $\varepsilon > 0$ such that
\begin{equation}
G + \varepsilon M M^\top + \frac{1}{\varepsilon} N^\top N \prec 0.
\label{eq:Petersen}
\end{equation}
\end{lem}
%
Note that Petersen's Lemma makes use of the spectral norm $\|\Delta\|=\max_i\lambda_i^{1/2}(\Delta\Delta^T)$, where $\lambda_i(A)$ are the eigenvalues of $A$.
Petersen's Lemma converts the problem of verifying sign-definiteness of~\eqref{eq:ineqpetersen} to a line search in the scalar parameter $\varepsilon>0$.
Note that $\Delta$ is only required to be norm-bounded, and thus Petersen's Lemma also applies to time-varying uncertainties so long as $\|\Delta(t)\|\le1$.
 Alternative proofs and generalizations of Petersen's Lemma can be found in \cite{Khlebnikov2008,Khlebnikov2014,BisoffiDePersisTesi22}.

\section{Local Stability Analysis and Stabilizing Controller Synthesis for Quadratic-Bilinear Systems}
\label{sec:synthesis}
Our objectives in this section are two-fold: (i)~determine an ellipsoidal estimate  for the ROA of the quadratic system~\eqref{eq:quad}, and (ii)~determine a static linear state feedback control that stabilizes the QB system~\eqref{eq:qb} within an ellipsoid centered about the origin.
Recall that the QB system~\eqref{eq:qb} reduces to the quadratic system~\eqref{eq:quad} when $u=0$.
As such, we will establish our main controller synthesis result to address~(ii), then proceed to state our main analysis result to address~(i) as a corollary.

Consider the QB system~\eqref{eq:qb} with the input determined by a static linear state feedback law as 
\begin{equation}
u = K x, \quad K \in \mathbb{R}^{m \times n}
\label{Controller}
\end{equation}
with \( m < n \).
Here, we seek a control gain $K$ that guarantees the closed-loop dynamics will be stable within an ellipsoid centered about the origin.
To this end, express the associated closed-loop dynamics of \eqref{eq:qb} with control~\eqref{Controller} as
\begin{equation}
    \dot{x} = (A + B K) x + \sum_{i=1}^n \begin{bmatrix} H_i & D_i \end{bmatrix}
    \begin{bmatrix} 
        x e_i^\top & 0 \\ 
        0 & x e_i^\top 
    \end{bmatrix}
    \begin{bmatrix} 
        I \\ 
        \tilde{K} 
    \end{bmatrix} x
    \label{eq:quad_feedback}
\end{equation}
where we have defined
\begin{align}
    D_i &:= 0 \quad \text{for } i = m+1, \dots, n, \\
    \tilde{K} &:= \begin{bmatrix}
        K \\
        0
    \end{bmatrix} \in \mathbb{R}^{n \times n}.
\end{align}
Next, introduce the quadratic Lyapunov function
\begin{equation}
    V(x) = x^\top Q x, \quad Q \succ 0.
    \label{eq:lyapunov_function}
\end{equation}
An expression for $\dot{V}(x)$ along trajectories of the quadratic system can be obtained by combining \eqref{eq:lyapunov_function} and \eqref{eq:quad_feedback} as

\begin{align}
    \dot{V}(x) &=  x\trans  Q \dot{x} + \dot{x}\trans  Q x \nonumber \\
    & = x\trans \left( Q(A+BK) + (A+BK)\trans Q   + Q\sum_{i=1}^n 
      \begin{bmatrix}
          H_i & D_i
      \end{bmatrix}
      \begin{bmatrix}
          xe_i\trans  & 0 \\
          0 & xe_i\trans 
      \end{bmatrix}
      \begin{bmatrix}
          I \\ \tilde{K} 
      \end{bmatrix}  + \sum_{i=1}^n 
      \begin{bmatrix}
           I & \tilde{K}\trans  
      \end{bmatrix}
      \begin{bmatrix}
          xe_i\trans  & 0 \\
          0 & xe_i\trans 
      \end{bmatrix}\trans 
      \begin{bmatrix}
          H_i\trans  \\ 
          D_i\trans 
      \end{bmatrix}Q\right)x.
  \label{eq:VdotCL}
\end{align}
For \( \dot{V}(x) < 0 \), we require
\begin{align*}
    & & Q(A+BK) + (A+BK)\trans Q   + Q\sum_{i=1}^n 
      \begin{bmatrix}
          H_i & D_i
      \end{bmatrix}
      \begin{bmatrix}
          xe_i\trans  & 0 \\
          0 & xe_i\trans 
      \end{bmatrix}
      \begin{bmatrix}
        I \\  \tilde{K}  
      \end{bmatrix}  + \sum_{i=1}^n 
      \begin{bmatrix}
         I & \tilde{K}\trans  
      \end{bmatrix}
      \begin{bmatrix}
          xe_i\trans  & 0 \\
          0 & xe_i\trans 
      \end{bmatrix}\trans 
      \begin{bmatrix}
          H_i\trans  \\ 
          D_i\trans 
      \end{bmatrix}Q  \prec 0.
\end{align*}
Pre- and post-multiplying the inequality above by the matrix \( P = Q^{-1} \succ 0 \) yields the equivalent condition
\begin{align}
    & (A+BK) P + P(A+BK)\trans   + \sum_{i=1}^n 
      \begin{bmatrix}
          H_i & D_i
      \end{bmatrix}
      \begin{bmatrix}
          xe_i\trans  & 0 \\
          0 & xe_i\trans 
      \end{bmatrix}
      \begin{bmatrix}
         P \\ \tilde{K} P 
      \end{bmatrix}  + \sum_{i=1}^n 
      \begin{bmatrix}
          P & P \tilde{K}\trans  
      \end{bmatrix}
      \begin{bmatrix}
          xe_i\trans  & 0 \\
          0 & xe_i\trans 
      \end{bmatrix}\trans 
      \begin{bmatrix}
          H_i\trans  \\ 
          D_i\trans 
      \end{bmatrix}  \prec 0.
    \label{LMi3}
\end{align}

This final expression is in a form conducive to exploiting Petersen's Lemma.
We are now ready to state our main synthesis result as Theorem~\ref{thm:main_synthesis}.

\begin{theorem}
    \label{thm:main_synthesis}
    Let $\varepsilon>0$ be given.
    If $P=P\trans\succ0$ and $Y\in\mathbb{R}^{m \times n}$ satisfy the LMI 
\begin{align}
\left[
\begin{array}{c|c}
    A P + P A^\top + B Y + Y^\top B^\top + \varepsilon \sum_{i=1}^{n} H_i P H_i^\top + \varepsilon \sum_{j=1}^{m} D_j P D_j^\top &\quad P \quad \begin{bmatrix} Y \\ 0 \end{bmatrix} \\
    \hline 
    P \\ \begin{bmatrix}
        Y^\top & 0
    \end{bmatrix} & -\varepsilon I
\end{array}
\right] 
\prec 0,
\label{eq:sdp_problem}
\end{align}
then the linear state feedback gain
\[
K = YP^{-1} 
\]
stabilizes the QB system~\eqref{eq:qb} inside the ellipsoid
\[
\hat{\mathcal{E}} = \{ x \in \mathbb{R}^n : x\trans  P^{-1} x \leq 1 \},
\]
and the quadratic form
\[
V(x) = x\trans  P^{-1} x
\]
is a Lyapunov function for the closed-loop system inside the stabilizability ellipsoid \(\hat{\mathcal{E}}\).
\end{theorem}

\begin{pf}
We proceed to establish equivalence with the conditions of Lemma~\ref{lemma:roahat} for the closed-loop QB system~\eqref{eq:quad_feedback}, beginning with conditions 2(a) and 2(b) then ending with condition 1.

Condition 2(a) is satisfied by construction of the Lyapunov function~\eqref{eq:lyapunov_function}.

For condition 2(b), $\dot{V}(0)=0$ by construction as seen in~\eqref{eq:VdotCL}.
To establish $\dot{V}(x)<0$ for all $x\in\ehat\backslash\{0\}$, we will make use of Petersen's Lemma.
First, let $\Delta =
    \begin{bmatrix}
          I_2 \otimes  e_1x\trans  P^{-\frac{1}{2}},&
         I_2 \otimes  e_2x\trans P^{-\frac{1}{2}}, & 
        \cdots, & 
         I_2 \otimes  e_nx\trans P^{-\frac{1}{2}}
    \end{bmatrix}\trans $, where $I_2\in\mathbb{R}^{2\times 2}$ is the identity matrix.
Accordingly, $\Delta\trans \Delta = \sum_{i=1}^n \left(I_2 \otimes  e_ix\trans  P^{-\frac{1}{2}}\right)\left(I_2 \otimes   P^{-\frac{1}{2}}xe_i\trans \right)=\sum_{i=1}^n (I_2)\otimes\left(e_ix\trans P^{-1}xe_i\trans \right)=(I_2)\otimes\left(x\trans P^{-1}x\right)$, and thus $\|\Delta\|_2\le1$ is equivalent to $x\trans P^{-1}x\le1.$
Next, invoke Lemma~\ref{lemma:petersen} (Petersen's Lemma) with  $G = (A+BK)P + P(A+BK)\trans $, $M  = \begin{bmatrix}
    D_1 P^{\frac{1}{2}} \quad H_1 P^{\frac{1}{2}} & 
    D_2 P^{\frac{1}{2}} \quad H_2 P^{\frac{1}{2}} & 
    \cdots & 
    D_n P^{\frac{1}{2}} \quad H_n P^{\frac{1}{2}}
\end{bmatrix}$, and $N = \begin{bmatrix}
        P\tilde{K}\trans &
        P
    \end{bmatrix}\trans $.
It follows that \eqref{LMi3} holds for all $x\in\ehat$
if and only if there exists a scalar $\varepsilon > 0$ such that $G + \varepsilon M M^\top + \frac{1}{\varepsilon} N^\top N \prec 0$.
For $\epsilon>0$ given, defining $Y:=KP$, applying the Schur Complement Lemma, and noting that $D_j=0$ for $j=m+1,\dots,n$ yields 
the desired LMI condition~\eqref{eq:sdp_problem}.
The control gain $K=YP^{-1}$ can be uniquely determined since $P\succ0$.

Finally, condition 1 requires that $\ehat$ be an invariant set, which is satisfied by virtue of the fact that $\ehat$ is defined by the Lyapunov function sublevel set $V(x)=x\trans P^{-1}x\le1$ on which $\dot{V}(x)<0$. \qed

\end{pf}

Our main analysis result follows as Corollary~\ref{thm:main_analysis}.

\begin{corollary}
\label{thm:main_analysis}
    Let \( \varepsilon > 0 \) be given. If \( P=P\trans\succ 0 \) is a solution to the LMI
    \begin{equation}
            \begin{bmatrix}
        AP + PA^\top + \varepsilon \sum_{i=1}^{n} H_i P H_i^\top  & P \\
        P & -\varepsilon I
    \end{bmatrix} \prec 0,
    \label{eq:lmi}
    \end{equation}
     then
    \begin{equation}        
\label{eq:ellipsoid}    \ehat = \left\{ x \in \mathbb{R}^n \ \bigg| \ x^\top P^{-1} x \leq 1 \right\}
    \end{equation}
    is an ellipsoidal approximation of the ROA for the quadratic system \eqref{eq:quad}, and  the quadratic form 
\begin{equation}
    V(x) = x^\top P^{-1} x
\end{equation}
    serves as a Lyapunov function for the system inside \( \ehat \).
\end{corollary}

\begin{pf}
    The result follows from Theorem 1 with $B=0$, $D_j=0$, $K=0$, $Y=0$. \qed
\end{pf}
For a given \(\varepsilon > 0\),  conditions \eqref{eq:sdp_problem} and \eqref{eq:lmi} are both LMI feasibility conditions. 
As such, Theorem~\ref{thm:main_synthesis} and Corollary~\ref{thm:main_analysis} define families of ellipsoidal approximations over which we can optimize.  
It is possible to optimize over these families to obtain desirable properties for the approximation, usually maximizing the size of the ellipsoidal set in some manner: e.g.,~maximizing the volume or the leading principal axis of $\ehat$.
Here, we seek ellipsoidal approximations for which the sum of the squares of the principal axes are maximized; i.e.,~for local stability analysis, given $\varepsilon>0$
    \begin{equation}    
    \max_{P\succ0} \quad \text{trace}(P) \quad \text{subject to~\eqref{eq:lmi}, } 
    \label{cor:trace}
    \end{equation}
    and for local stabilization, given $\varepsilon>0$ 
   \begin{equation}    
    \max_{P\succ0,Y} \quad \text{trace}(P) \quad \text{subject to~\eqref{eq:sdp_problem} } 
    \label{cor:largest_stabilizability_ellipsoid}
    \end{equation}
with the associated optimal controller gain determined as $K=YP^{-1}$.

The optimizations in~\eqref{cor:trace} and~\eqref{cor:largest_stabilizability_ellipsoid} yield ellipsoidal approximations $\ehat\subseteq\roa$ of the ROA and ROS, respectively, for fixed values of $\varepsilon>0$.
In either case, a bisection can be performed over $\varepsilon$ to obtain the maximum such ellipsoidal approximation $\ehat^*$ with corresponding $\varepsilon=\varepsilon^*>0$.

\begin{remark}
\label{rem:union}
In light of the discussions above on optimality, it is important to note the following:
Even if an optimal ellipsoidal approximation can be found, this approximation may not encapsulate the complete ROA or ROS, which in general may not even be ellipsoidal.
It is possible to improve the ROA or ROS approximation by considering multiple ellipsoidal approximations~\cite{Chesi2007}.
Consider a collection of ellipsoidal approximations $\ehat_\ell$ with $\ell=1,\dots,L$ determined from pairs $(P_\ell,\varepsilon_\ell)$---not necessarily optimal---for which~\eqref{eq:lmi} holds.
Since $\ehat_\ell\subseteq\roa$, a more complete approximation of the ROA or ROS can be constructed as the union of these ellipsoidal approximations, i.e.,
\begin{equation}
        \roahat=\bigcup_{\ell=1}^L \ehat_\ell \subseteq \roa.
        \label{eq:union}
\end{equation}
Furthermore, the ROA or ROS approximation $\roahat$ constructed as in~\eqref{eq:union} will not necessarily be ellipsoidal nor even convex.

 \end{remark}

\begin{remark}
\label{Remark2}
The formulations thus far pertain to asymptotic stability and stabilization of the origin, but can be extended to consider exponential stability and stabilization with a specified minimum decay rate $\alpha>0$.
To do so, modify~\eqref{eq:VdotCL} to reflect the condition $\dot{V}(x) \le -\alpha V(x)$ and combine with the fact that
\begin{equation}
G + M\Delta N + N\trans \Delta\trans M\trans \le G+\varepsilon MM\trans  +\frac1\varepsilon N\trans N     \quad \forall\varepsilon>0
\end{equation}
for all admissible $\Delta$ in Petersen's Lemma~\cite{Khlebnikov2016}.
Taking $G$, $M$, $\Delta$, and $N$ as specified in the proof of Theorem~\ref{thm:main_synthesis} yields
      \begin{equation}
        PA\trans  + AP + BY + Y^\top B^\top + \varepsilon \sum_{i=1}^n H_i P H_i\trans  +  \varepsilon \sum_{j=1}^{m} D_j P D_j^\top+\frac{1}{\varepsilon} \begin{bmatrix}
            P & \begin{bmatrix}
                Y \\ 0
            \end{bmatrix}
        \end{bmatrix}\begin{bmatrix}
            P \\ \begin{bmatrix}
                Y^\top & 0
            \end{bmatrix}
        \end{bmatrix} \preceq -\alpha P.
        \label{eq:modifyLMI1}
    \end{equation} 

        %
        By the Schur Complement Lemma, we arrive at an LMI condition for exponential stabilization with minimum decay rate $\alpha>0$ within an ellipsoid centered about the origin as,
        \begin{align}
\left[
\begin{array}{c|c}
    A P + P A^\top + B Y + Y^\top B^\top + \varepsilon \sum_{i=1}^{n} H_i P H_i^\top + \varepsilon \sum_{j=1}^{m} D_j P D_j^\top + \alpha P &\quad P \quad \begin{bmatrix} Y \\ 0 \end{bmatrix} \\
    \hline 
    P \\ \begin{bmatrix}
        Y^\top & 0
    \end{bmatrix} & -\varepsilon I
\end{array}
\right] 
\preceq 0.
\label{eq:sdp_problem1}
\end{align}
The analog for determining the region of exponential stability follows by taking  $B = 0$, $D_j = 0$, $K = 0$, and $Y = 0$ as,
    \begin{equation}
        \label{eq:sdp_problem2}
        \begin{bmatrix}
            A P + P A\trans  + \varepsilon \sum_{i=1}^{n} H_i P H_i^\top + \alpha P & P \\
            P & -\varepsilon I
        \end{bmatrix}
        \preceq 0.
    \end{equation}

\end{remark}

\section{Illustrative Examples}
\label{sec:analysis_ex}
We apply the proposed analysis and synthesis methods on model systems
 studied within the context of local stability analysis and stabilization in prior studies.
Results for all examples reported here were computed in Matlab using the convex optimization modeling language \texttt{cvx}~\cite{cvx} in conjunction with the optimization solver \texttt{MOSEK}~\cite{mosek}. All examples are run on a 2022 MacBook Air with an Apple M2 chip, 8\,GB of memory, and macOS~Sonoma~14.5.

\subsection{Local Stability Analysis: Two-State Quadratic System from~\cite{amato2006region}}
Consider the quadratic system given by
\begin{align}
    \dot{x}_1 &= -50 x_1 - 16 x_2 + 13.8 x_1 x_2, \nonumber \\
    \dot{x}_2 &= 13 x_1 - 9 x_2 + 5.5 x_1 x_2. 
    \label{2state}
\end{align}
Local stability analysis of this system was originally studied in~\cite{amato2006region} and more recently in~\cite{LiaoSeilerHemati22}.
The system dynamics in~\eqref{2state} can be transcribed into the format of~\eqref{eq:quad} as
\begin{align}
x=\begin{bmatrix}x_1\\x_2\end{bmatrix},\quad    A &= \begin{bmatrix}
        -50 & -16 \\ 
        13 & -9
    \end{bmatrix}, \quad
    H = \begin{bmatrix}
        0 & 6.9 & 6.9 & 0 \\ 
        0 & 2.75 & 2.75 & 0
    \end{bmatrix}. \label{eq:AH_matrices}
\end{align}
Note that $A$ is Hurwitz and $H$ is symmetric as defined in~\eqref{eq:Hsymm}.

\begin{figure}[htb]
    \centering
    \subfigure[]{%
        \includegraphics[width=0.45\linewidth]{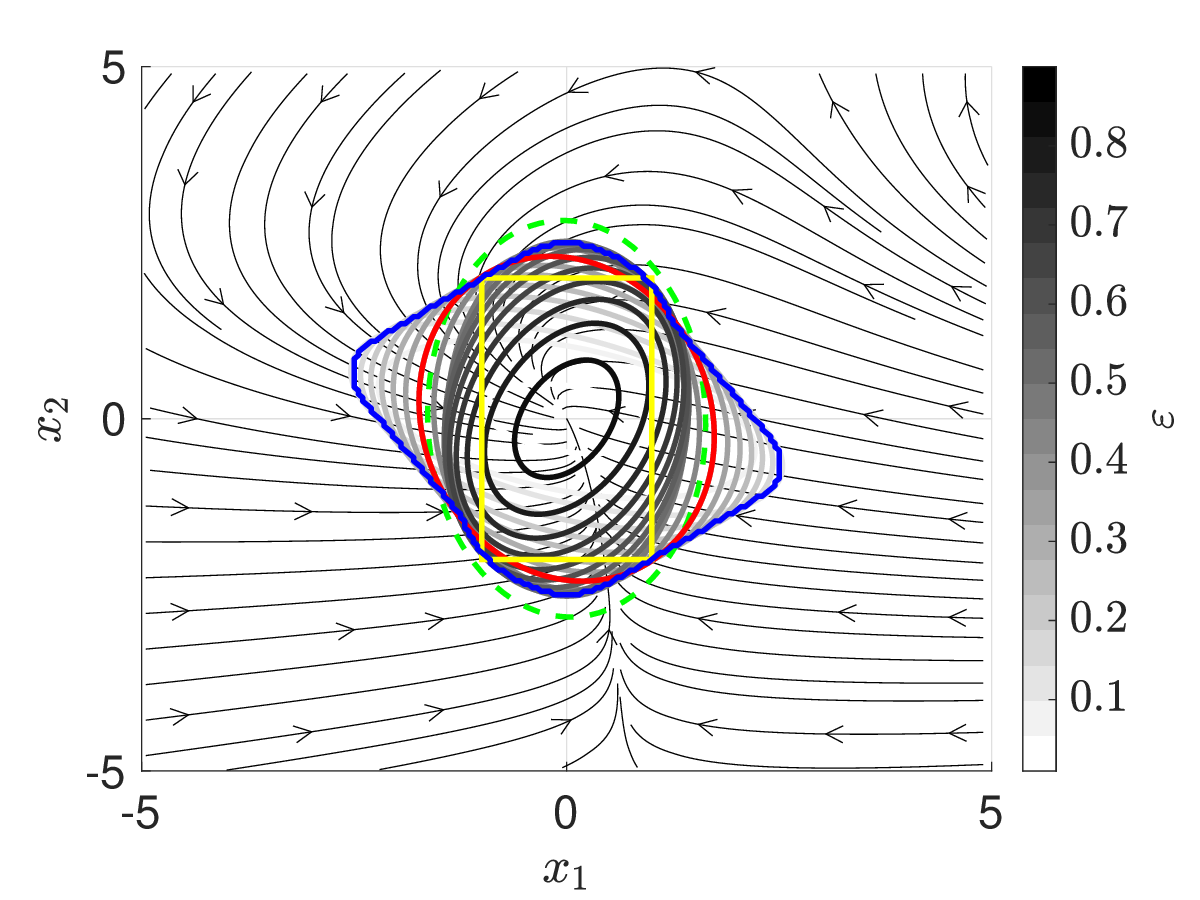}%
        \label{fig:state2_a}
    }
    \quad
    \subfigure[]{%
        \includegraphics[width=0.45\linewidth]{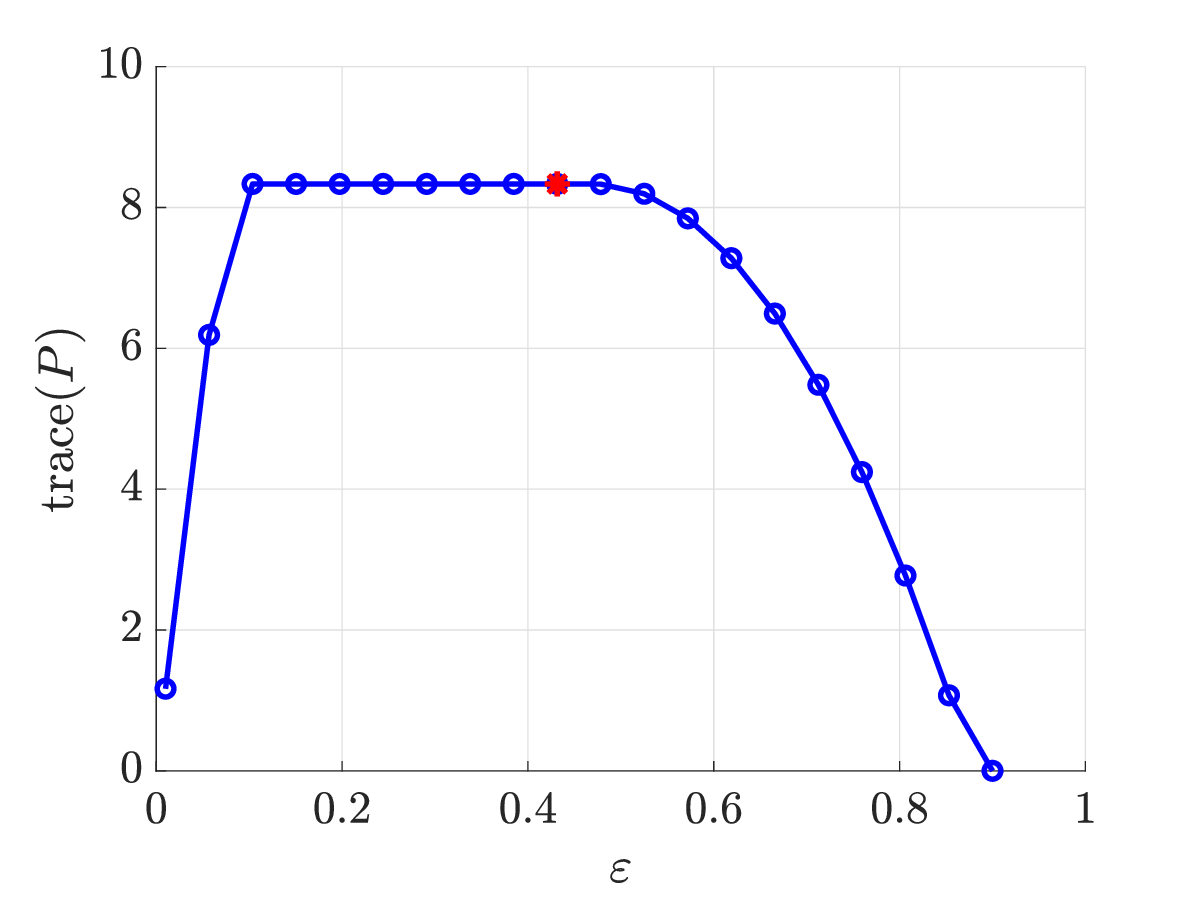}%
        \label{fig:state2_b}
    }
    \caption{ROA estimates for the 2-state example from~\cite{amato2006region}. 
    (a)~ROA estimates are overlaid on the system's phase portrait.
    Ellipsoidal  estimates from~\eqref{cor:trace}  are plotted in grayscale, and the ROA estimate resulting from their union in blue.
        The polytope and associated ellipsoidal ROA estimate from~\cite{amato2006region} are plotted in yellow and green, respectively).
   (b)~Quantitative results obtained from~\eqref{cor:trace} gridded on~$\varepsilon$.
    The red ellipsoid and the red marker in (a) and (b), respectively, correspond to the maximum trace ellipsoid over the set computed.
    }
    \label{fig:phase_plane}
\end{figure}

Optimal ellipsoidal approximations for the ROA of~\eqref{2state} are found by solving~\eqref{cor:trace} over a uniformly spaced grid of $\varepsilon\in[0.01,0.8]$ with 20 points. 
%
%
The resulting ellipsoidal approximations are overlaid on a phase portrait for the system in Figure~\ref{fig:state2_a} with the associated $\varepsilon$ and $\mathrm{trace}(P)$ shown in Figure~\ref{fig:state2_b}. 
The ellipsoidal approximation with the maximum trace from this set and the associated $\varepsilon$ are highlighted in red in Figure~\ref{fig:state2_b}.
For all ROA approximations shown, we can visually confirm that all trajectories entering the ROA remain in the ROA and ultimately decay to the origin.
 An interesting feature in this example is that $\mathrm{trace}(P)$ remains relatively constant for $\varepsilon \in [0.09, 0.53]$. The associated ellipsoidal approximations thus have roughly the same size but are oriented differently in the state space. Consequently, the union of these ellipsoids yields a non-ellipsoidal approximation of the ROA (highlighted in blue in Figure~\ref{fig:state2_a}). 
 Note that for $\varepsilon$ values outside this range, the size of the ellipsoid becomes significantly smaller.
 Further, the LMI condition~\eqref{eq:lmi} was found to be infeasible outside the range of $\varepsilon\in[0.01,0.8]$ considered.

A direct comparison is made in Figure~\ref{fig:phase_plane} with the analysis results reported by Amato et al. in~\cite{amato2006region} (yellow and green).
The approach in~\cite{amato2006region} identifies the polytope $[-1,\,1] \times [-2,\,2]$ (yellow) as an ROA, which is achieved by circumscribing the polytope within a particular ellipsoidal approximation of the ROA (green).
Note this circumscribing ellipsoid (green) has {$\mathrm{trace}(P)=10.604$}, whereas the largest ellipsoidal approximation from our analysis (red) has {$\mathrm{trace}(P)=8.3347$}. 
It is interesting to note that our non-ellipsoidal ROA approximation (blue) nearly circumscribes the same polytope reported in~\cite{amato2006region}, and covers an area of $15.9825$ units$^{2}$---compared with $14.4639$ units$^{2}$ for the (green) ellipsoid from~\cite{amato2006region} and $12.8340$ units$^{2}$ for the (red) ellipsoid from our trace maximization.
Note that the LMI condition for certifying that a polytope constitutes an ROA approximation---as in~\cite{amato2006region}---grows exponentially with the state dimension, i.e.,~$\sim\mathcal{O}(2^n)$.
This rapid growth in problem size results from the number of parameters required to define a polytope in $n$-dimensional space.
In contrast, our LMI condition in~\eqref{eq:lmi} grows quadratically with the state dimension, i.e.,~$\sim\mathcal{O}(n^2)$.

\begin{figure}[h]
   \centering
    \includegraphics[width=0.45\columnwidth]{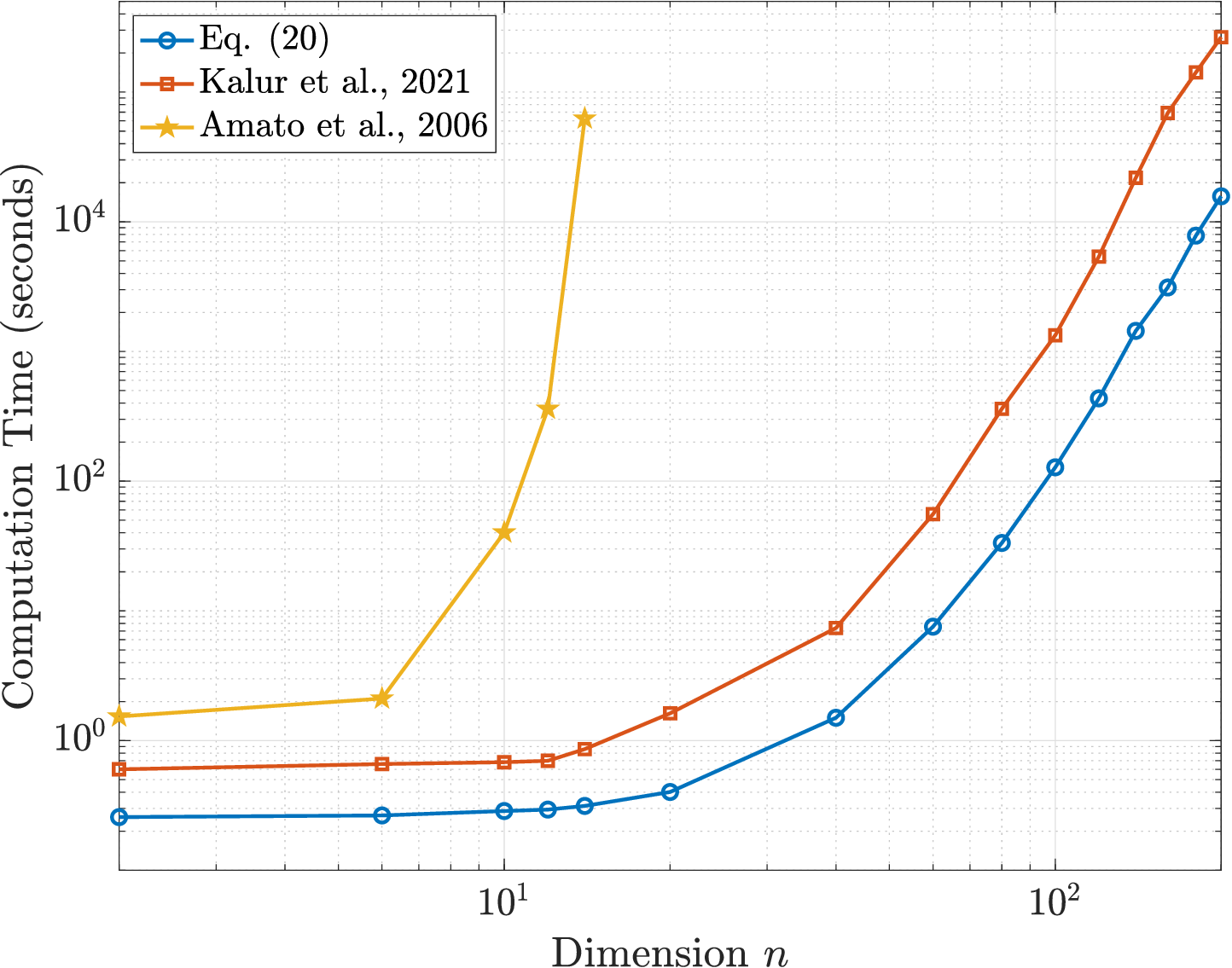} 
\caption{An empirical comparison of computation time versus state dimension $n$ for~\eqref{cor:trace}, the  quadratic constraint method of Kalur et al.~\cite{KalurEtAl21}, and the polytopic estimation method of Amato et al.~\cite{amato2006region} based on ``stacking'' the 2-state example from~\cite{amato2006region}.}
   \label{computation_time}
\end{figure}

Next, we empirically evaluate the computational complexity of evaluating~\eqref{cor:trace}.
To do so, we construct higher-dimensional systems by ``stacking'' the system in~\eqref{2state}.
Figure~\ref{computation_time} compares the computation of solving~\eqref{cor:trace}  with the quadratic constraint~(QC) analysis approach from~\cite{KalurEtAl21} and the polytopic stability certification method from~\cite{amato2006region}.
Note that computation times for the polytopic stability certification method is reported up to $n=14$, which corresponds to roughly the same wall-clock time required to solve~\eqref{cor:trace} or the QC method with $n=200$.
Fitting a power law expression for the compute time $t\sim n^k$ to the results for~\eqref{cor:trace} and the QC method for $n\ge40$ yields \(k={5.91}\) and \(k={6.76}\), respectively.
An exponential fit  to the full set of polytopic stability certification results yields a compute-time scaling of \(t\sim2.26^n\).
A quantitative comparison of ROA estimates with the QC method of~\cite{KalurEtAl21} will be made in the next section.

\subsection{Local Stability Analysis: 9-State Model of a Sinusoidally-Forced Shear Flow~\cite{Moehlis2004}}
Consider the reduced-order model of a sinusoidally-forced shear flow given by
\begin{equation}
\dot{x} = A(Re)x + H(x\otimes x)
\end{equation}
where the scalar parameter $Re>0$ is the Reynolds number and $x\in \mathbb{R}^9$ is the state vector.
This low-order model was originally formulated in~\cite{Moehlis2004} based on a Fourier mode representation of the incompressible Navier-Stokes equations and has been found to exhibit salient features of transitional and turbulent flows.
The model has been used in a variety of recent works investigating local stability analysis methods aimed at predicting transition to turbulence~\cite{heidePRF2024,KalurEtAl21,KalurSeilerHemati21,Liu2020,goulart2012}.
For an in-depth discussion of the model and associated coefficients in \( A \) and \( H \), we point the reader to the original paper~\cite{Moehlis2004}.

\begin{figure}[htb]
    \centering
    \subfigure[Trace(P) vs $\varepsilon$ for Differnt Reynold Numbers ]{%
        \includegraphics[width=0.45\linewidth]{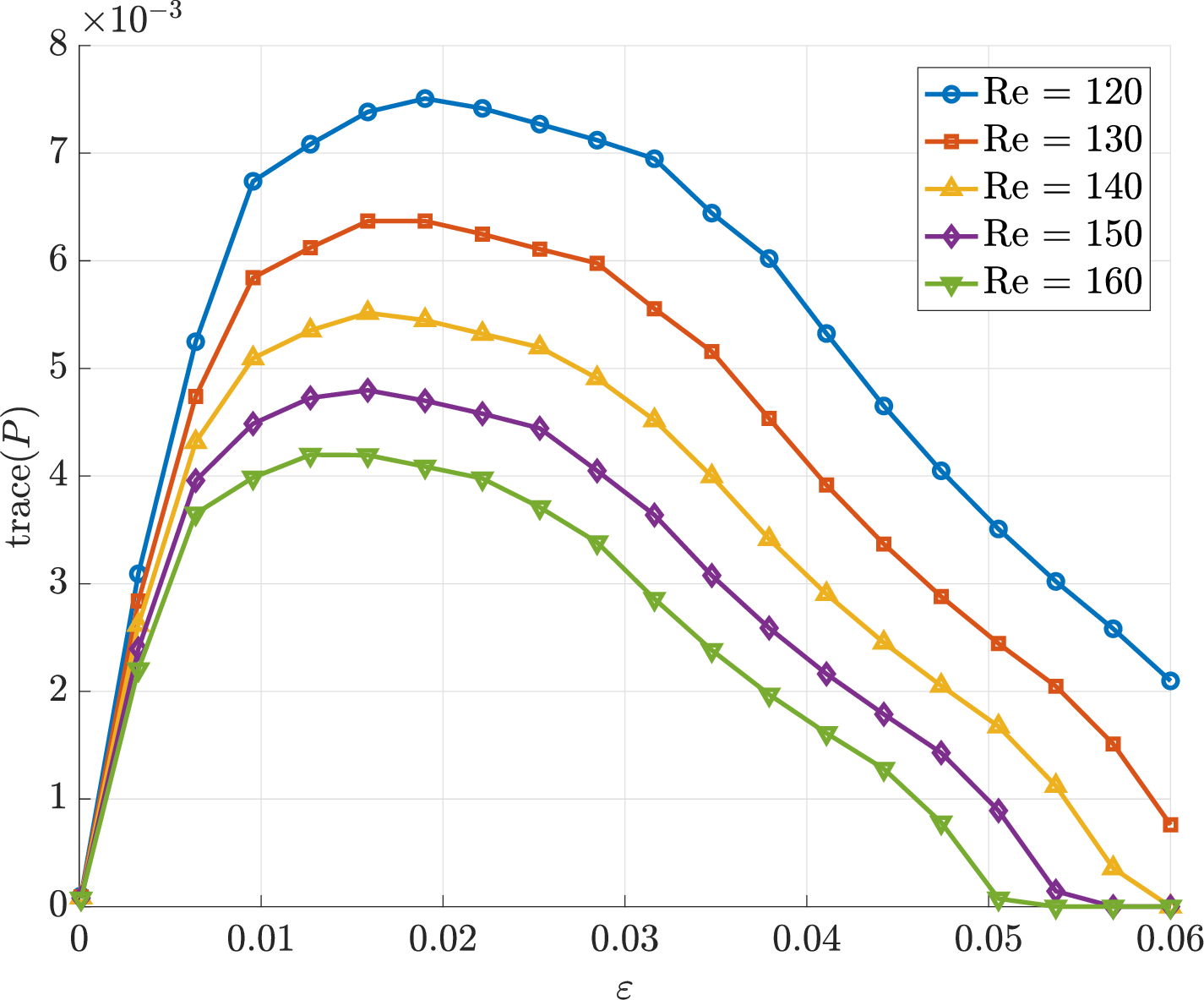}%
        \label{fig:state9_a}
    }
    \quad
    \subfigure[Trace(P) over Re]{%
        \includegraphics[width=0.45\linewidth]{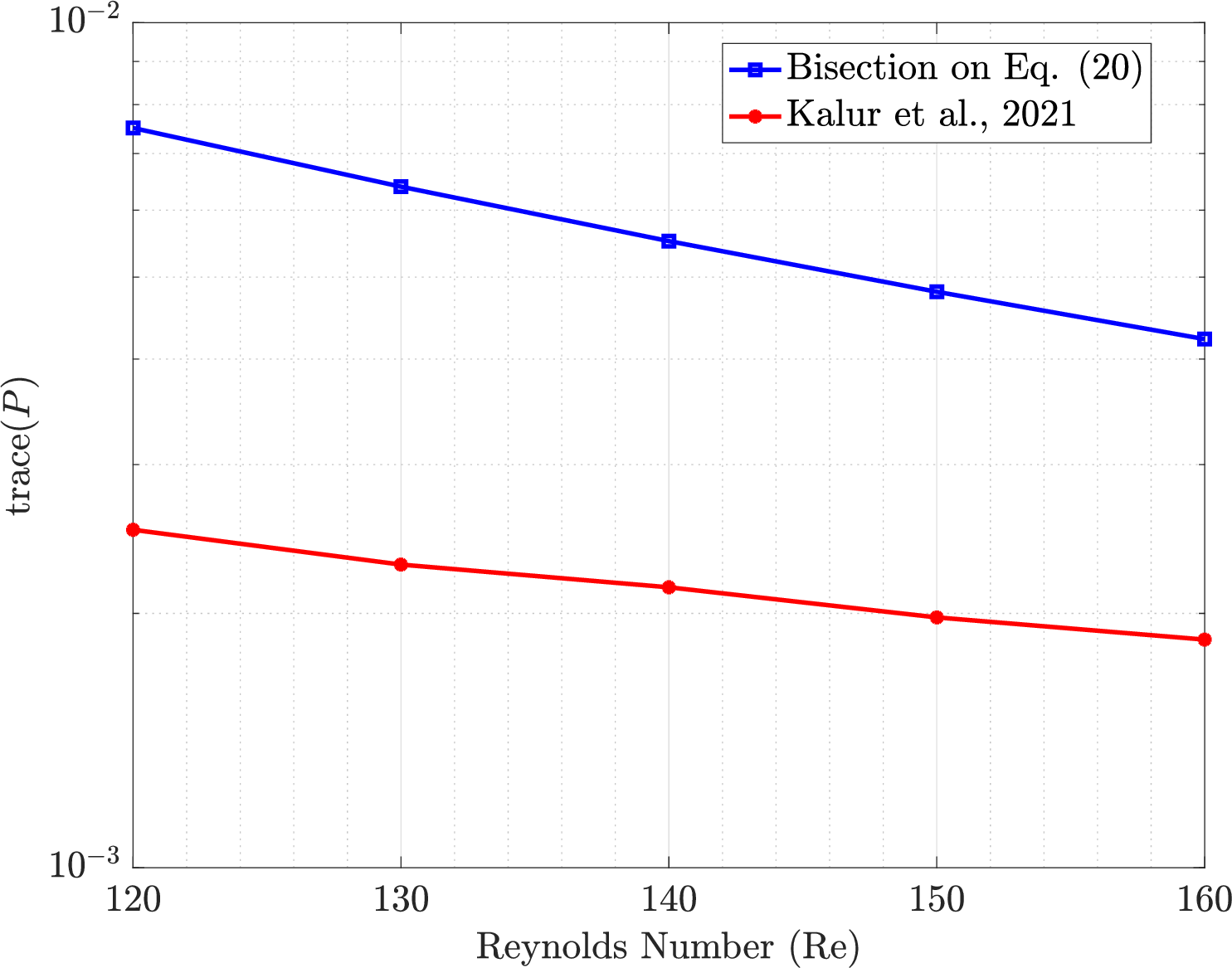}%
        \label{fig:state9_b}
    }
    \caption{ROA estimates for the 9-state model from~\cite{Moehlis2004} versus Reynolds number~($Re$). (a)~Results from applying~\eqref{cor:trace} over a grid of~$\varepsilon$ for different $Re$. (b)~Comparison of the maximum ellipsoidal ROA estimate from~\eqref{cor:trace} determined via bisection (blue) is compared with ellipsoidal ROA estimates based on the quadratic constraint method of Kalur et al.~\cite{KalurEtAl21} (red).} 
    \label{fig:state9}
\end{figure}

We apply~\eqref{cor:trace} over a range of $Re$ (see Figure~\ref{fig:state9}).
For each $Re$, we used a uniformly spaced grid of \( \varepsilon\in[10^{-3},1] \) with 50 distinct values; representative results for $Re=\{120,130,140,150, 160\}$ are shown in Figure~\ref{fig:state9_a}. 
A bisection procedure is employed to determine the maximum value of $\mathrm{trace}(P)$ and the associated optimal $\varepsilon$ for each Reynolds number ($Re$). 
The results are presented in Figure~\ref{fig:state9_b}, where we compare our approach (blue) with that of~\cite{KalurEtAl21} (red). 
Bisection on~\eqref{cor:trace} yields a larger (in trace) estimate for the region of attraction compared to the QC approach of~\cite{KalurEtAl21}.
All these findings are consistent with the underlying physics: as $Re$ increases, the region of attraction (ROA) shrinks, indicating a growing susceptibility to instability and transition.

\subsection{Stabilizing Controller Synthesis: Three-state Quadratic-Bilinear System from \cite{Amato2009}}

Consider stabilization of the QB system from~\cite{Amato2009}
\begin{align}
    \dot{x}_1 &= -1.7 x_1 + 1.7 x_2 +  x_1 u_1-x_2u_2 + 0.8u_1 + 3.2u_2, \nonumber \\
    \dot{x}_2 &= 1.37 x_1 -  x_2 - 0.7x_3 -x_2x_3+1.1u_1+0.2u_2,  \nonumber \\
    \dot{x}_3 &= 0.7 x_1 +  x_2 -1.6x_3+ 0.2 x_1 x_2 + x_1u_1+0.5x_2u_1 - x_2u_2-0.1x_3u_2+7.5u_1+0.6u_2.
    \label{3state}
\end{align}
Transcribing into the form of \eqref{eq:qb}, we have
\begin{align}
A &= \begin{bmatrix}
        -1.7 & 1.7 & 0 \\ 
        1.37 & -1 & -0.7 \\
        0.7 & 1 & -1.6
    \end{bmatrix}, \quad
    D_1 = \begin{bmatrix}
        0 & -1 & 0\\0 & 0 & 0\\1 & \frac{1}{2} & 0
    \end{bmatrix}, \quad D_2 = \begin{bmatrix}
        1 & 0 & 0\\0 & 0 & 0\\0 & -1 & 0.1
    \end{bmatrix}, \quad B = \begin{bmatrix}
        0.8 & 3.2\\ 1.1 & 0.2 \\ 7.5 & 0.6
    \end{bmatrix} \nonumber \\
   & H_1 = \begin{bmatrix}
        0 & 0 & 0 \\ 0 & 0 & 0\\0 & 0.1 & 0
    \end{bmatrix}, \quad  H_2 = \begin{bmatrix}
        0 & 0 & 0 \\ 0 & 0 & -\frac{1}{2}\\0.1 & 0 & 0
    \end{bmatrix}, \quad  H_3 = \begin{bmatrix}
        0 & 0 & 0 \\ 0 & -\frac{1}{2} & 0\\0 & 0 & 0
    \end{bmatrix}. \label{eq:AH_matrices}
\end{align}
Note that $A$ is Hurwitz and $H$ is symmetric as defined in~\eqref{eq:Hsymm}.
Although the system is already locally stable, the example was used to demonstrate local stabilization within a polytope in~\cite{Amato2009}.
We use this example to make a direct comparison with the results from~\cite{Amato2009} here.

\begin{figure}[htb]
    \centering
    \subfigure[]{%
        \includegraphics[width=0.45\linewidth]{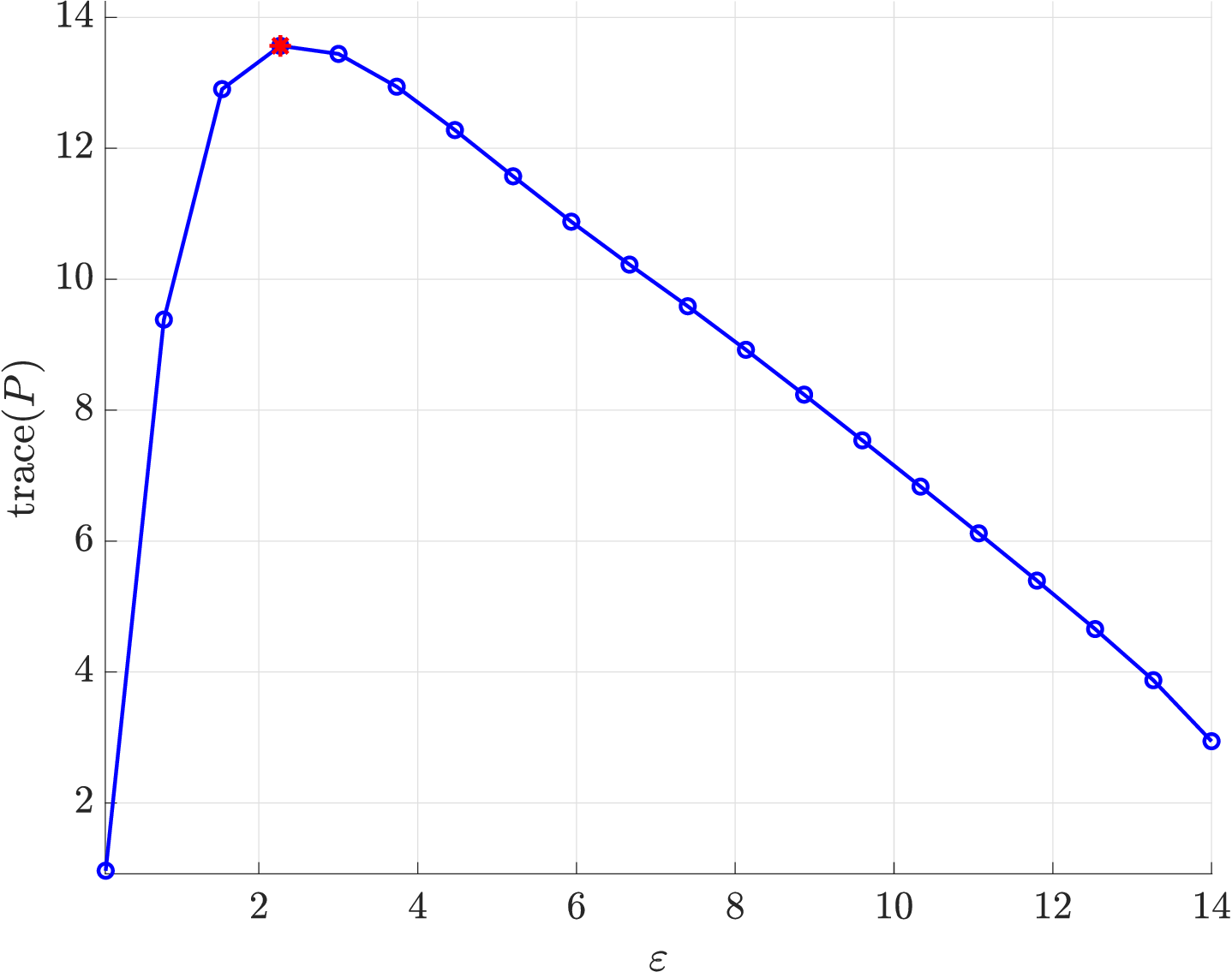}%
        \label{fig:petersen1}
    }
    \subfigure[]{%
        \includegraphics[width=0.45\linewidth]{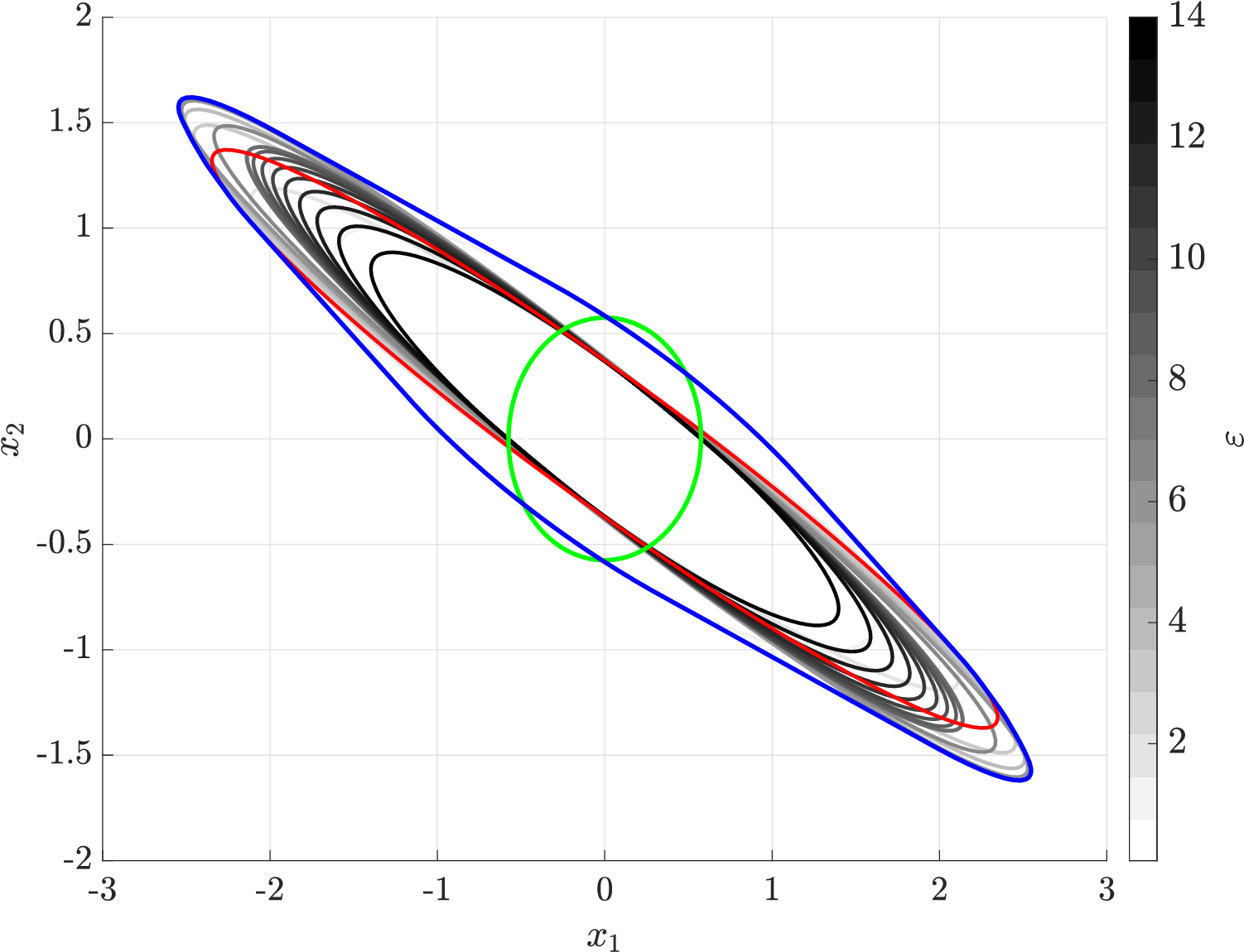}%
        \label{fig:petersen2}
    }
    \subfigure[]{%
        \includegraphics[width=0.45\linewidth]{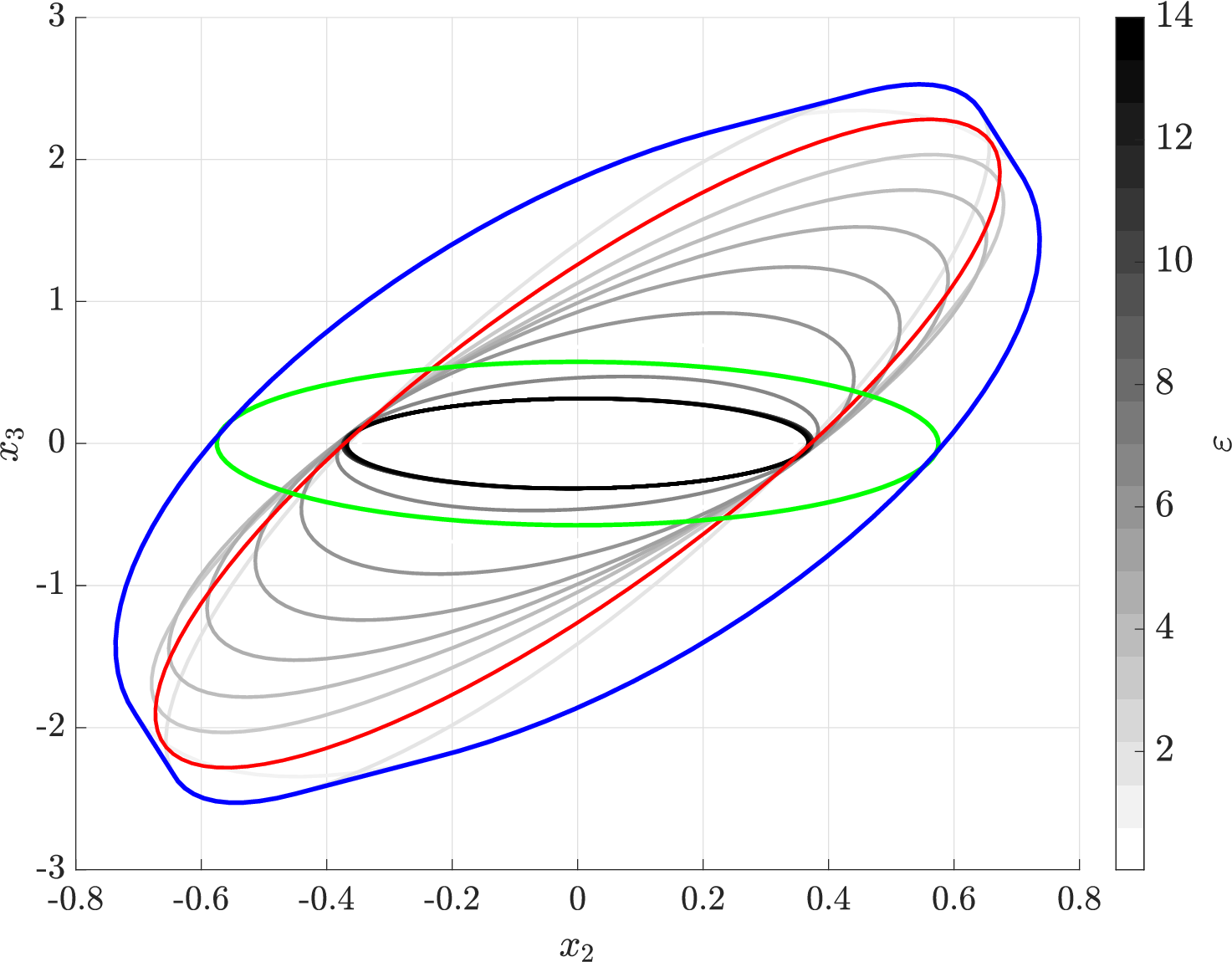}%
        \label{fig:petersen3}
    }
    \subfigure[]{%
        \includegraphics[width=0.45\linewidth]{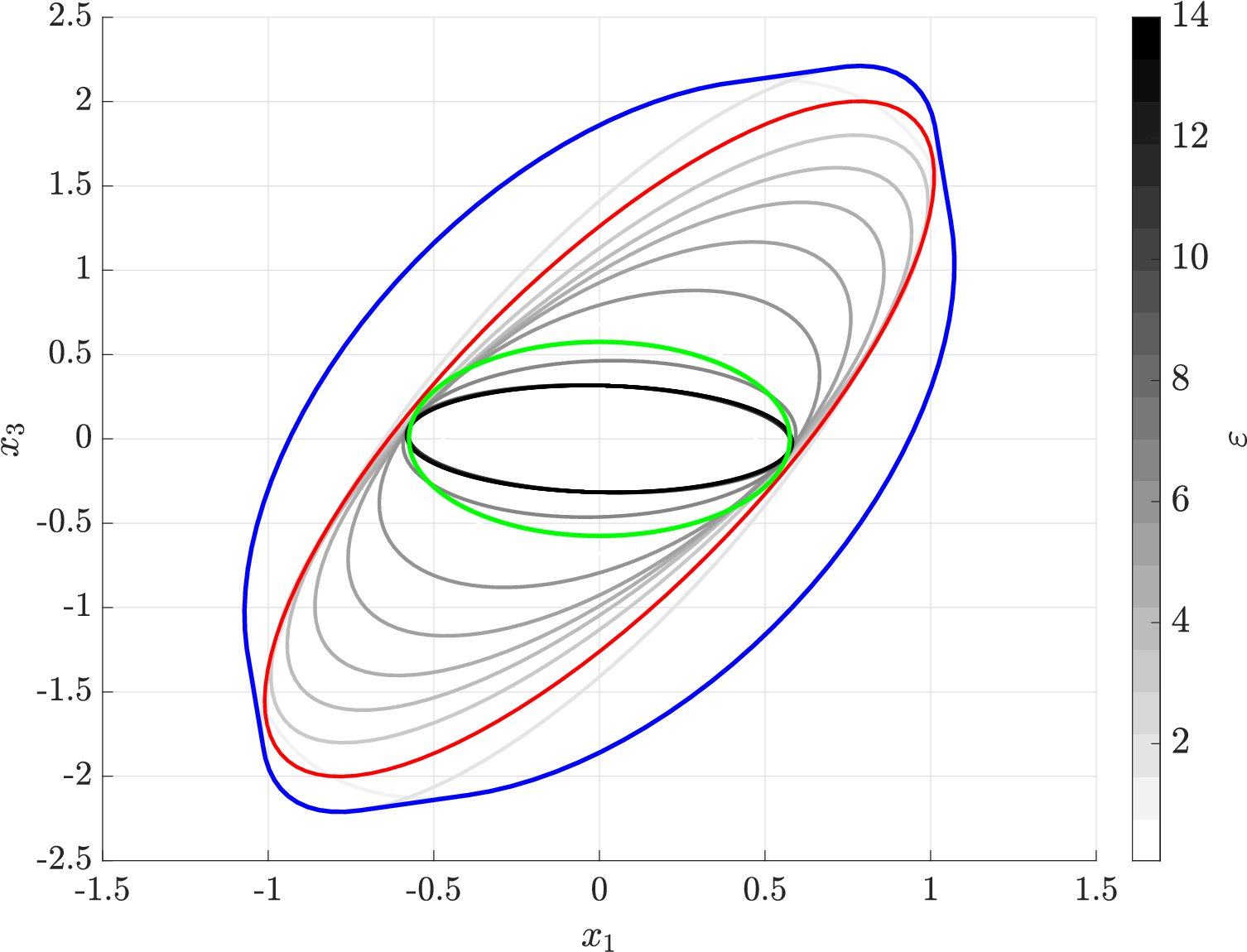}%
        \label{fig:petersen4}
    }
    \caption{ROS estimates for the 3-state example from~\cite{Amato2009}.
    (a) Results from applying~\eqref{cor:largest_stabilizability_ellipsoid} over a grid of $\varepsilon$, with the maximum estimate highlighted in red. (b)–(d) Cross-sections of ellipsoidal estimates from~\eqref{cor:largest_stabilizability_ellipsoid} (grayscale), the ROS estimate determined from their union (blue), and the largest ellipsoidal estimate based on the polytopic ROS method of~\cite{Amato2009}.}
    \label{fig:petersen_results}
\end{figure}

Optimal ellipsoidal approximations for the region of stabilizability~(ROS) of \eqref{3state} are obtained by solving \eqref{cor:largest_stabilizability_ellipsoid} over a uniformly spaced grid for $\varepsilon \in [0.01,14]$ with 20 grid points.
Note that the LMI condition \eqref{eq:sdp_problem} was found to be infeasible for $\varepsilon$ outside this range.
Figure~\ref{fig:petersen1} reports the resulting $\mathrm{trace}(P)$ for each value of $\varepsilon$ considered, with the associated ellipsoidal estimates for the ROS shown in Figures~\ref{fig:petersen2}--\ref{fig:petersen4} as planar slices in the $x_1$--$x_2$, $x_2$--$x_3$, and $x_1$--$x_3$ planes, respectively.  
The largest (in trace) ROS estimate among the set is highlighted in red. 
The union of ellipsoidal estimates yields a non-ellipsoidal approximation of the ROS (blue), corresponding to a region in which there exists a linear state feedback control law~\eqref{Controller} capable of stabilizing the QB system~\eqref{3state}.
A direct comparison is made with the synthesis approach proposed in \cite{amato2007state} (green). The method was applied to stabilize the QB system~\eqref{3state} within a prescribed polytope  $[-1,\,1] \times [-1,\,1] \times [-1,\,1]$.
The largest (in trace) ellipsoid estimate for the ROS we were able to find is highlighted in green.
This ellipsoid has {$\mathrm{trace}(P)=0.9927$}, which is  lower---sometimes by an order of magnitude---than all the ellipsoidal estimates found by solving~\eqref{cor:largest_stabilizability_ellipsoid} (see Figure~\ref{fig:petersen1}).

\section{Conclusions}
We have presented an approach for analyzing the local stability and designing stabilizing controllers for quadratic-bilinear~(QB) systems using Petersen's Lemma. By utilizing quadratic Lyapunov functions and formulating the stability and control synthesis conditions as linear matrix inequalities (LMIs), we have provided computationally efficient methods suitable for high-dimensional systems.
We have shown that the analysis and synthesis methods scale to systems with hundreds of state variables using standard convex optimization software and without resorting to specialized implementations.
Although we have formulated the analysis and synthesis methods for continuous-time QB systems, the extension to discrete-time systems is straightforward.
Extensions to state estimation and dynamic compensation for QB systems will be the focus of future work.

\bibliographystyle{plain}        

\bibliography{main}           

\end{document}